\documentclass[useAMS, usenatbib]{mn2e}

\usepackage{epsfig}

\usepackage{amssymb}

\usepackage{multirow}

\usepackage{enumitem}

\usepackage{ulem}

\usepackage{rotating}
\usepackage{natbib}
\usepackage{latexsym}
\bibpunct{(}{)}{;}{a}{}{,}
\usepackage{url}

\begin{document}

\newcommand{\ledd}{%
$L_{Edd}$}

\newcommand{\IGR}{IGR~J18245--2452}

\newcommand{\Msun}{M$_\odot$}

\def\rem#1{{\bf #1}}
\def\hide#1{}

\def \aj {AJ}
\def \mnras {MNRAS}
\def \apj {ApJ}
\def \apjs {ApJS}
\def \apjl {ApJL}
\def \aap {A\&A}
\def \aapr {A\&ARv}
\def \nat {Nature}
\def \araa {ARAA}
\def \pasp {PASP}
\def \aaps {AAPS}
\def \prd {PhRvD}
\def \apss {ApSS}

\newcommand{\specialcell}[2][c]{%
  \begin{tabular}[#1]{@{}c@{}}#2\end{tabular}}

\author[Linares et al.]{
\parbox[t]{\textwidth}{
\raggedright 
Manuel Linares$^{1,2}$\thanks{Linares@iac.es},
Arash Bahramian$^{3}$,
Craig Heinke$^{3}$,
Rudy Wijnands$^{4}$,
Alessandro Patruno$^{5,6}$,
Diego Altamirano$^{4,7}$,
Jeroen Homan$^{8}$,
Slavko Bogdanov$^{9}$,
David Pooley$^{10,11}$
}
\vspace*{6pt}\\
$^1$ Instituto de Astrof{\'i}sica de Canarias, c/ V{\'i}a L{\'a}ctea
s/n, E-38205 La Laguna, Tenerife, Spain\\
$^2$ Universidad de La Laguna, Departamento de Astrof{\'i}sica,
E-38206 La Laguna, Tenerife, Spain\\
$^3$ Department of Physics, University of Alberta, CCIS 4-183, Edmonton, AB
T6G 2E1, Canada\\
$^4$ Astronomical Institute ``Anton Pannekoek'', University of
Amsterdam, Science Park 904, 1098XH Amsterdam, The Netherlands\\
$^5$ Leiden Observatory, Leiden University, PO Box 9513, 2300RA
Leiden, The Netherlands\\
$^6$ ASTRON, The Netherlands Institute for Radio Astronomy, Postbus 2,
7990-AA Dwingeloo, The Netherlands\\
$^7$ Physics \& Astronomy, University of Southampton, Southampton,
Hampshire SO17 1BJ, UK\\
$^8$ Kavli Institute for Astrophysics and Space Research,
Massachusetts Institute of Technology, 70 Vassar Street, Cambridge, MA
02139, USA\\
$^9$ Columbia Astrophysics Laboratory, Columbia University, 550 West
120th Street, New York, NY 10027, USA\\
$^{10}$ Department of Physics, Sam Houston State University, Huntsville,
TX, USA\\
$^{11}$ Eureka Scientific, Austin, TX, USA\\
 }

\title[M28: from outburst to quiescence]{The neutron star transient
and millisecond pulsar in M28: from sub-luminous accretion to
rotation-powered quiescence}

\date{Accepted 2013 November 6. Received 2013 October 14; in original form 2013 September 4.}

\maketitle

\begin{abstract}

The X-ray transient IGR~J18245--2452 in the globular cluster M28
contains the first neutron star (NS) seen to switch between
rotation-powered and accretion-powered pulsations.
We analyse its 2013 March-April 25~d-long outburst as
observed by {\it Swift}, which had a peak bolometric luminosity of
$\sim$6\% of the Eddington limit (L$_{Edd}$), and give detailed
properties of the thermonuclear burst observed on 2013 April 7.
We also present a detailed analysis of new and archival {\it Chandra}
data, which we use to study quiescent emission from
IGR~J18245--2452 between 2002 and 2013.
Together, these observations cover almost five orders of magnitude in
X-ray luminosity (L$_X$, 0.5--10~keV).
The {\it Swift} spectrum softens during the outburst decay (photon
index $\Gamma$ from 1.3 above L$_X$/L$_{Edd}$$=$10$^{-2}$ to
$\sim$2.5 at L$_X$/L$_{Edd}$$=$10$^{-4}$), similar to other NS and
black hole (BH) transients.
At even lower luminosities, L$_X$/L$_{Edd}$=[10$^{-4}$--10$^{-6}$], deep
{\it Chandra} observations reveal hard ($\Gamma$=1--1.5), purely
non-thermal and highly variable X-ray emission in quiescence.
We therefore find evidence for a spectral transition at
L$_X$/L$_{Edd}$$\sim$10$^{-4}$, where the X-ray spectral softening
observed during the outburst decline turns into hardening as the
source goes to quiescence.
Furthermore, we find a striking variability pattern in the 2008 {\it
Chandra} light curves: rapid switches between a high-luminosity
``active'' state (L$_X$$\simeq$3.9$\times$10$^{33}$~erg~s$^{-1}$) and
a low-luminosity ``passive'' state
(L$_X$$\simeq$5.6$\times$10$^{32}$~erg~s$^{-1}$), with no detectable
spectral change.
We put our results in the context of low luminosity accretion flows
around compact objects and X-ray emission from millisecond radio
pulsars.
Finally, we discuss possible origins for the observed mode switches in
quiescence, and explore a scenario where they are caused by fast
transitions between the magnetospheric accretion and pulsar wind shock
emission regimes.

\end{abstract}


\section{Introduction}
\label{sec:intro}

Globular clusters have proved to be excellent locations for the study
of low-mass X-ray binaries (LMXBs). This is due in part to the
dynamical production of LMXBs through stellar interactions, which
enhances the number of LMXBs per unit stellar mass by a factor of
$\sim$100 over the rest of the Galaxy \citep[e.g.,][]{Clark75}, and
makes clusters excellent targets to identify and study LMXBs.  The
knowledge of the companion's age and metallicity, and the tight
constraints on the distance and the absorbing column density ($N_H$)
provided by optical observations, also assist in studies of LMXBs in
clusters \citep[e.g.][]{Kuulkers03}.

Eighteen luminous (0.5--10~keV luminosity L$_X>10^{35}$ erg~s$^{-1}$)
LMXBs have been reported to date in 15 Galactic globular clusters
(including IGR~J18245--2452). In all cases the accreting compact
object has been identified as a neutron star (NS), mostly thanks to
the detection of thermonuclear bursts (in 16 systems) or
accretion-powered pulsations (in 4 systems). About half are
transients, detectable in outburst with wide-field X-ray monitors for
a few days to months \citep[see Table 5 in][and references
therein]{Bahramian13}. Most LMXBs in globular clusters are usually in
deep quiescence ($L_X<10^{33}$ erg~s$^{-1}$), undergoing little or no
accretion.  An estimated total of $\sim$200 quiescent LMXBs exist in
the Galactic globular cluster system, identified through deep
observations with the high-resolution Chandra satellite
\citep{Pooley03,Heinke05}. Quiescent NS-LMXBs typically produce
blackbody-like thermal emission from their surfaces at
$L_X\sim10^{32-33}$ erg~s$^{-1}$ \citep{Rutledge00}, although a
substantial subset of quiescent NS-LMXB spectra are dominated by
non-thermal emission of unknown origin, perhaps produced by continued
low-level accretion or a pulsar wind shock
\citep{Campana98,Wijnands05c}. The thermal NS surface emission is
understood to be due to the release of heat deposited in the deep
crust during accretion outbursts \citep{Brown98,Rutledge02}.

The globular cluster M28 (NGC 6626) is home to 12 known millisecond
radio pulsars \citep[MRPs,][]{Begin06,Bogdanov11}.
Numerous other faint X-ray sources in the cluster were identified with
deep {\it Chandra}-ACIS observations in 2002 and 2008, including a
relatively bright source (S26) with a soft spectrum indicative
of a quiescent NS-LMXB \citep{Becker03, Servillat12}.
An unusual, subluminous ($\sim$0.02 of the Eddington luminosity) X-ray
burst was detected from the direction of M28's core by ASCA, during a
time when no LMXB was obviously active in the cluster, suggesting
low-level accretion in quiescence \citep{Gotthelf97}.

A new X-ray transient was discovered by {\it INTEGRAL} on 2013 March
28, from a position consistent with M28, and named IGR~J18245--2452
\citep{Eckert13}. Using {\it Swift}-XRT observations, \citet{Heinke13}
confirmed the association of the new transient with the core of M28.
\citet{Romano13} and \citet{Barthelmy13} reported analyses
of the X-ray light curve and spectrum of \IGR{} near the outburst peak,
which triggered the {\it Swift}-BAT on three occasions
\citep{Romano13}. A type I (thermonuclear) X-ray burst from
\IGR{} was detected with the {\it Swift}-XRT on 2013 April 07
\citep{Papitto13,Linares13}, identifying the source as a
NS-LMXB. \citet{Pavan13} reported the radio detection and
sub-arcsecond position of \IGR{} in outburst, on 2013 April 05, which
is consistent with only one source from \citet{Becker03}: S23.  Using
the improved radio location a variable likely optical counterpart to
\IGR{} was found in archival {\it HST} observations
\citep{Pallanca13,Cohn13}. A {\it Chandra} observation taken on 2013
April 28 \citep{Homan13} showed no new bright sources in the core of
M28, and S23 at a luminosity higher than that measured in 2002.

\citet{Papitto13b} discovered 254~Hz X-ray pulsations during two {\it
XMM} observations taken on 2013 Apr. 3 \& 13, and identified the new
transient NS-LMXB with a previously known MRP in an 11~hr orbit:
PSR~J1824--2452I \citep{Begin06}. MRPs \citep{Backer82} have long been
thought to be the evolutionary end point of NS-LMXBs
\citep{Alpar82}. The discovery of accretion-powered millisecond
pulsars \citep[AMPs;][]{Wijnands98} provided strong evidence for such
evolutionary link. While other indirect evidence has been presented
\citep{Burderi03,Archibald09}, \citet{Papitto13b} found \IGR{} to be the
first system directly observed to switch between MRP and AMP
phases at different times. \IGR{} was detected again as an MRP after the
2013 outburst \citep{Papitto13c}.
Burst oscillations were found in the April 07 burst observed by {\it
  Swift}, at the 254~Hz spin frequency
\citep{Patruno13,Papitto13b,Riggio13}.

We present a detailed analysis of the 2013 March-April outburst of
\IGR{} as observed by {\it Swift}, as well as the thermonuclear burst
that led to its NS-LMXB classification (Sec.~\ref{sec:burst}). We also
study the quiescent properties of \IGR{}, using deep {\it Chandra}
observations of M28, and we find that it features a hard, purely
non-thermal and highly-variable quiescent spectrum. We find evidence
for a spectral transition around $L_X/L_{Edd} \sim$10$^{-4}$
(Sec.~\ref{sec:ob2q}), and a striking variability pattern in the
longest {\it Chandra} observations of \IGR{} in quiescence
(Sec.~\ref{sec:qui}). We discuss our results in
Section~\ref{sec:discussion}, in the context of low-luminosity
accretion flows, quiescent emission from NS-LMXBs and X-ray emission
from MRPs.

\begin{figure*}
  \begin{center}
  \resizebox{2.0\columnwidth}{!}{\rotatebox{0}{\includegraphics[]{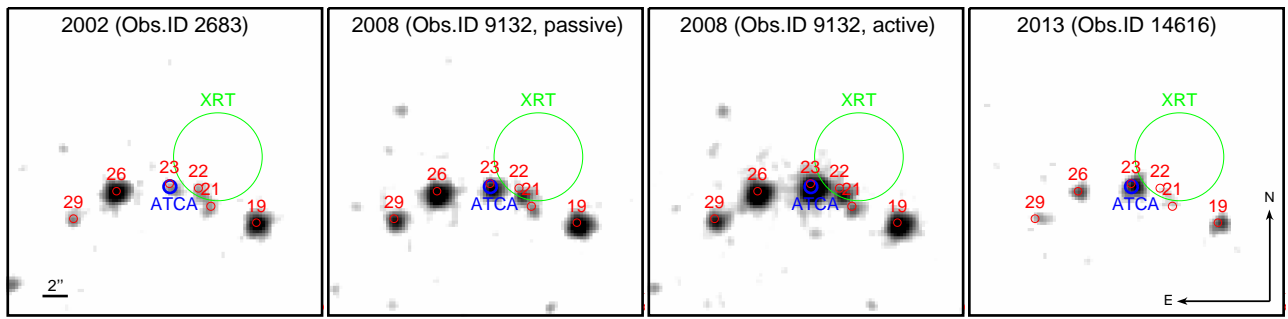}}}
  \caption{
{\it Chandra}-ACIS-S archival observations of the central parts of M28
in four epochs (from left to right): 2002, 2008 during the passive
state, 2008 during the active state and 2013 after the outburst of
\IGR{} (see text for details). Image comparison showing the sources from
\citet[][red circles]{Becker03}, the {\it Swift}-XRT \citep[][green
circle]{Romano13} and {\it ATCA} \citep[][blue circle]{Pavan13}
locations of the 2013 outburst from IGR~J18245--2452. Only source 23
from \citet[][S23]{Becker03}, marginally consistent with the {\it
Swift}-XRT location (90\% c.l. uncertainty of 3.5$\arcsec$), lies
inside the ATCA (90\% c.l.)  0.5$\arcsec$ error circle (which is also
consistent with the {\it Chandra}-HRC position reported by
\citealt{Papitto13b}).
} %
    \label{fig:qimg}
 \end{center}
\end{figure*}

\section{Data Analysis}
\label{sec:data}

We give in this Section a detailed explanation of the data reduction
and analysis procedures. All luminosities, radiated energies and radii
in this work use 5.5~kpc, the distance reported by \citet[][2010
revision]{Harris96} based on HST photometry of M28/NGC~6626
\citep{Testa01}. When normalizing luminosities to the Eddington limit,
we use $L_{Edd}$=2.5$\times 10^{38}$~erg~s$^{-1}$. Unless otherwise
noted, uncertainties are quoted at the 1$\sigma$ confidence
level and fluxes and luminosities are given in the 0.5--10~keV
band. Since hard states emit a sizeable fraction of their total
luminosity outside the 0.5--10~keV band covered by the {\it
Swift}-XRT, a bolometric correction ($f_{bol}$=3) is applied when
estimating mass accretion rates \citep[see, e.g.,][]{intZand07}.

\subsection{Swift}
\label{sec:swift}

We analysed all {\it Swift}-XRT observations of M28 available in
August 2013, totalling 28 observations and about 37 ksec of on-source
time, including pre-planned (target numbers 90442, 32785, 32787) and
triggered (triggers 2336, 2369, 2370) observations. All products were
extracted using FTOOLS (v. 6.12). The only {\it Swift} observation
taken before the discovery outburst of IGR~J18245-2452 (on 2010-06-24)
shows a blend of very faint unresolved sources in the core of M28
\citep[][see below]{Becker03}, and was not included in the rest of the
analysis.

One persistent and one background spectrum were extracted per
observation (after running {\it xrtpipeline}, v. 0.12.6) using pointed
data in windowed timing (WT) or photon counting (PC) mode (taking the
longest exposure when both modes were present in the same
observation). A circular extraction region with a 30-pixel radius
centered on the source position \citep{Romano13} was used for most
observations; an annular region with 5- and 40-pixel radii was used in
the brightest PC mode observations in order to correct for
pile-up. The X-ray burst was excluded to obtain the persistent
spectrum of the corresponding observation. We created ancilliary
response files using {\it xrtmkarf} (v. 0.6.0) and an exposure map for
each event file using {\it xrtexpomap} (v. 0.2.7), thereby correcting
for vignetting and the XRT's point spread function (PSF). After
grouping the spectra to a minumum of 15 counts per bin, we fitted them
in the 0.5--10~keV band within {\it Xspec}
\citep[v. 12.7.1;][]{Arnaud96} using the latest response matrices from
the calibration database (0to12s6\_20070901v011 and
0to2s6\_20070901v012 for PC and WT mode, respectively) and a simple
absorbed power law model.

{\it Swift}-XRT cannot resolve the numerous faint ($L_X \lesssim
10^{33}$ erg~s$^{-1}$) X-ray sources often present in the center of
globular clusters. In the case of M28, twelve sources have been
detected with {\it Chandra} within the 0.24' core radius
\citep{Becker03}, which is similar to {\it Swift}-XRT's half-power
diameter, 0.3'. The standard background subtraction method, which uses
a source free region outside the globular cluster's core, does not
take into account the contribution from nearby faint X-ray sources in
the center of M28, which we refer to as ``unresolved faint-source
background'' (UFB). We investigated whether or not this additional
background component affects the flux and spectral parameters from
\IGR{} measured with the XRT, as follows. We extracted a spectrum from
the 2002 and 2008 {\it Chandra}-ACIS observations of M28
(Sec.~\ref{sec:chandra}) using the same ($\sim$1.2' radius) region
used for {\it Swift}-XRT extraction but excluding our source of
interest, \IGR{}. The resulting UFB spectrum does not change between the
2002 and 2008 observations and can be fitted with an absorbed power
law model with photon index 1.96$\pm$0.06 (90\% c.l.;
$N_H=2\times10^{21}$ cm$^{-2}$), which yields a 0.5--10~keV luminosity
of $\sim 4.5 \times 10^{33}$ erg~s$^{-1}$ and an XRT count rate of
$\sim$1.8$\times 10^{-2}$~c~s$^{-1}$.

We then repeated the XRT spectral fits of the persistent emission,
adding a second power-law component to the model with parameters
frozen at the values found for the UFB spectrum \citep[and a
multiplicative constant fixed at 0.926 to take into account the
absolute flux calibration difference between ACIS-S3 and the
XRT,][]{Tsujimoto11}. Thus the final model used was phabs*(powerlaw +
constant*powerlaw), where the second term is kept fixed and accounts
for the UFB. We thereby fit only excess emission above the UFB, i.e.,
from our source of interest (the only known variable source within the
extraction region). 

For luminosities above $10^{34}$ erg~s$^{-1}$ we
find spectral parameters fully consistent with those found with the
standard background subtraction method. When \IGR{}'s luminosity drops
below $10^{34}$ erg~s$^{-1}$, however, its flux becomes comparable to
that of the UFB and its spectrum cannot be well constrained with {\it
Swift}-XRT. We therefore excluded those observations taken between Apr
19 and May 06, when \IGR{} was too faint to be disentangled from the
UFB. We also fitted together three observations taken on consecutive
days (Apr 15-17) in order to improve the spectral constraints on \IGR{}
at the faintest luminosities accessible to {\it Swift} ($\sim 4 \times
10^{34}$ erg~s$^{-1}$). Finally, we verified that fixing $N_H$ to the
values found with {\it Chandra} yields consistent results ($N_H \equiv
2.6\times10^{21}$ cm$^{-2}$ gives photon indices lower by 10\%, and a
similar softening with decreasing $L_X$, although the fits with this
model are worse than those using our preferred free N$_H$ model;
Sec.~\ref{sec:results}).

\begin{table*}
\center
\footnotesize
\caption{Observations and best-fit parameters for the {\it Chandra}-ACIS spectra of IGR~J18245--2452 in quiescence.}
\begin{minipage}{\textwidth}
\begin{tabular}{l c c c c c c c c}
\hline\hline
Epoch & Observation & Date & Exp. & Luminosity\footnote{Luminosity in the 0.5-10~keV band. Flux in the 0.5-10~keV band, absorbed/observed.}  & Flux$^a$ & $\Gamma$\footnote{Power law photon index (photon flux $\propto$ E$^{-\Gamma}$).} & N$_H$\footnote{Absorbing column density; frozen values are noted with a ``$\equiv$'' symbol.} & $\chi^2$/dof \\
 & IDs & (dd/m) & (ksec) & (erg~s$^{-1}$) & (erg~s$^{-1}$~cm$^{-2}$) & & (10$^{21}$ cm$^{-2}$) & \\
\hline
2002 & 268-4,5,3 & 04/7,04/8,11/9 & 41 &  [2.5$\pm$0.8]$\times 10^{32}$ & [6.6$\pm$2.0]$\times 10^{-14}$ & 0.8$\pm$0.3 & $\equiv$2.6 & 4.0/3 \\
2002\footnote{Fit to the same 2002 spectrum using Cash's C-statistic.} & 268-4,5,3 & 04/7,04/8,11/9 & 41 & [2.2$\pm$0.4]$\times 10^{32}$ & [5.4$\pm$1.0]$\times 10^{-14}$ & 1.2$\pm$0.2 & $\equiv$2.6 & 294.2/647 \\
2008 & 9132(Average) & 07/8 & 144 & [3.0$\pm$0.1]$\times 10^{33}$ & [7.2$\pm$0.3]$\times 10^{-13}$ & 1.44$\pm$0.05 & 2.6$\pm$0.2 & 257.3/249 \\
2008-A & 9132(Active) & 07/8 & 97 & [3.9$\pm$0.1]$\times 10^{33}$ & [9.2$\pm$0.4]$\times 10^{-13}$ & 1.51$\pm$0.04 & 2.9$\pm$0.2 & 33.3/27 \\
2008-P & 9132(Passive) & 07/8 & 47 & [5.6$\pm$1.0]$\times 10^{32}$ & [1.3$\pm$0.2]$\times 10^{-13}$ & 1.45$\pm$0.15 & 2.6$\pm$0.8 & 29.7/29 \\
2013 & 14616 & 28/4 & 15 & [3.8$\pm$0.7]$\times 10^{33}$ & [8.9$\pm$1.6]$\times 10^{-13}$ & 1.6$\pm$0.2 & $\equiv$2.6 & 9.5/15 \\
\hline\hline
\end{tabular}
\end{minipage}
\label{table:qspec}
\end{table*}

\begin{figure}
  \begin{center}
  \resizebox{1.0\columnwidth}{!}{\rotatebox{0}{\includegraphics[]{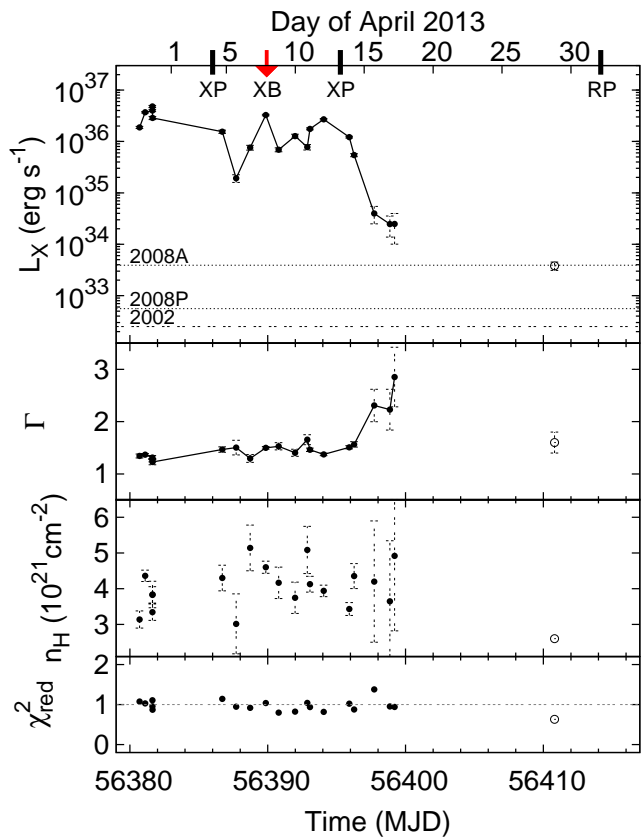}}}
  \caption{Luminosity (0.5--10 keV) and spectral evolution measured by
  {\it Swift} during the 2013 April outburst of
  IGR~J18245--2452. Thick lines across the top axis mark the times of
  the thermonuclear burst (red arrow labelled ``XB'';
  Fig.~\ref{fig:burst}), the two detections of millisecond X-ray
  pulsations reported by \citet[][XP]{Papitto13b} and the first
  detection of radio pulsations after the outburst found by
  \citet[][RP]{Papitto13c}. Open circles show data from our {\it
  Chandra} observation and horizontal lines on the top panel show our
  quiescent luminosity measurements from previous {\it Chandra}
  observations, as indicated (Table~\ref{table:qspec}).}
    \label{fig:ob}
 \end{center}
\end{figure}

For the burst time-resolved spectroscopy, we extracted contiguous
spectra in 2s steps (4s in the last stages of the burst tail) and used
a $\sim$100s-long interval as background, therefore assuming that the
persistent emission remains constant during the burst. As the burst
peak count rate was higher than 300 c/s, above the nominal pile-up
limit for WT mode, we used an annular extraction region with radii of
1 and 20 pixels centered on the XRT position \citep{Romano13}. We
fitted the resulting spectra as above, but using this time an absorbed
black body model with $N_H$ fixed at 4.4$\times$10$^{21}$~cm$^{-2}$,
the value reported from XRT spectroscopy of IGR~J18245--2452's
persistent emission \citep[][we also checked that leaving $N_H$ free
gives consistent results]{Romano13,Heinke13}.

\subsection{Chandra}
\label{sec:chandra}

We analysed all publicly available {\it Chandra}-ACIS-S observations
of M28, totalling 41~ksec in 2002 (July-August; obs. IDs 2684, 2685
and 2683; see \citealt{Becker03}) and 199~ksec in 2008 (August;
obs. IDs 9132 and 9133; see \citealt{Bogdanov11}). Following the
discovery of the new transient \IGR{}
\citep[][Sec.~\ref{sec:intro}]{Eckert13}, we obtained a 15~ksec {\it
Chandra} ToO observation of M28 on 2013-04-28 \citep[obs. ID
14616,][]{Homan13}, which we also analyse herein.
Table~\ref{table:qspec} shows a summary of the {\it Chandra}-ACIS
observations and spectra.
We also analysed the two publicly available {\it Chandra}-HRC-S
observations of M28, taken in 2002 (November, obs. 2797) and 2006
(May, obs. 6769). Even though no spectral information is available
in the HRC data, we extracted background-corrected count rates from
\IGR{} using a 1.6''-radius circular region, in order to constrain the
source quiescent luminosity at all possible times.
All products were extracted using tools and scripts from the latest
available CIAO\footnote{Chandra Interactive Analysis of Observations,
available at \url{http://cxc.harvard.edu/ciao/}} package
\citep[v. 4.5;][]{Fruscione06}.

Figure~\ref{fig:qimg} shows four smoothed {\it Chandra}-ACIS-S images
of M28's core taken in 2002, 2008 during the passive state, 2008
during the active state and 2013 after the outburst of \IGR{} (see
Table~\ref{table:qspec} for observation IDs and exposure times). See Sections
\ref{sec:qui} and \ref{sec:pulsar}, respectively, for a definition and
discussion of active/passive states. Superposed are the initial {\it
Swift}-XRT location of \IGR{} \citep[][green circle]{Romano13}, the
position of the {\it ATCA} radio counterpart found during its 2013
outburst \citep[][blue circle]{Pavan13} and the nearby faint X-ray
sources studied by \citet[][red circles]{Becker03}. Only source 23
from \citet{Becker03} is consistent with the ATCA position \citep[as
noted by][]{Pavan13,Papitto13b}, which is also consistent with the
{\it Chandra}-HRC location reported by \citet{Papitto13b}, and
marginally consistent with the {\it Swift}-XRT location \citep[][which
was not ``enhanced'' with UVOT attitude correction]{Romano13}. Hence
we confirm the association of the new transient source \IGR{} with
source 23 of \citet{Becker03}.

Source spectra were extracted from primary event files using circular
regions of 2--4 pixel radii and the script {\it specextract} (v. 10),
which creates the corresponding ancilliary response files corrected
for {\it Chandra}'s PSF. We grouped the resulting spectra to a minimum
of 15 counts per energy bin, and extracted the background spectrum
from a 10-pixel circular source-free region. In order to increase the
signal-to-noise ratio (S/N) in the 2002 spectrum of IGR~J18245--2452's
quiescent counterpart, we summed the spectra and event files from all
three observations using the CIAO scripts {\it combine\_spectra} and
{\it reproject\_obs}, respectively.

\section{Results}
\label{sec:results}

\subsection{Outburst evolution and thermonuclear burst}
\label{sec:burst}

Figure~\ref{fig:ob} shows an overview of the 2013 outburst from \IGR{}
including the evolution of the luminosity, photon index, and the
reported detection dates of the thermonuclear burst
\citep{Papitto13,Linares13}, X-ray pulsations \citep{Papitto13b} and
reappearance of radio pulsations \citep{Papitto13c}. A simple absorbed
power law model provides a good fit to the persistent
(accretion-powered) {\it Swift}-XRT spectra, and yields photon indices
in the range 1.2--2.8.
We find a maximum 0.5--10~keV outburst luminosity of
(4.8$\pm$0.2)$\times$ 10$^{36}$~erg~s$^{-1}$ ($\sim$6\% of $L_{Edd}$
assuming a bolometric correction factor $f_{bol}$=3) on 2013 March 30,
and a power law (photon) index of 1.26$\pm$0.03 on the same
date. These luminosities and photon indices, as well as the burst
behaviour (see below) and the presence of millisecond X-ray pulsations
\citep{Papitto13b}, are all consistent with the properties of the
so-called atoll sources \citep[a sub-class of low-luminosity
  NS-LMXBs,][]{HK89} in the hard state \citep{Linares09d}.

\begin{figure}
  \begin{center}
  \resizebox{1.0\columnwidth}{!}{\rotatebox{0}{\includegraphics[]{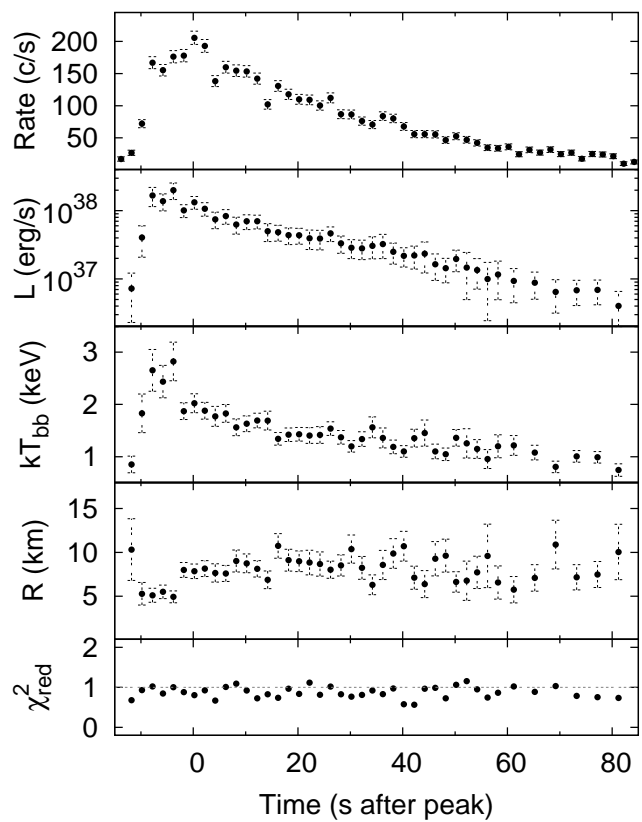}}}
  \caption{Net count rate (0.5--10~keV), bolometric luminosity and
  spectral evolution during the type-I (thermonuclear) X-ray burst
  from IGR~J18245--2452 detected by {\it Swift} on 2013-04-07
  \citep{Papitto13,Linares13}.}
    \label{fig:burst}
 \end{center}
\end{figure}

We present the main properties and spectral evolution of the
thermonuclear burst detected with {\it Swift}-XRT in
Table~\ref{table:burst} and Figure~\ref{fig:burst}, respectively. The
burst spectra are well fitted with a simple absorbed blackbody model
and show the unequivocal temperature drop (``cooling tail'') along the
flux decay, which is consistent with a simple exponential decay with
an e-folding time of 26.6$\pm$1.8~s. The count rate light curve shows
a $\sim$10~s-long plateau near its maximum with a possible double-peak
structure, both commonly seen in type I X-ray burst profiles. We do
not find spectral evidence of photospheric radius expansion around the
burst peak \citep[e.g.,][]{Kuulkers03}, and the bolometric luminosity
does not exceed 2$\times$10$^{38}$~erg~s$^{-1}$. The blackbody radii
that we measure during the burst (without color/redshift corrections)
are in the [5--10]~km range, also typical of thermonuclear bursts from
NSs.

\begin{table}
\center
\caption{Properties of the type I X-ray burst from IGR~J18245--2452 observed by {\it Swift} on 2013-04-07.}
\begin{minipage}{\textwidth}
\begin{tabular}{ l c}
\hline\hline
Peak time (of max. net rate)  (UTC) & 22:15:42 \\
Rise time (25\% to 90\% of peak rate)  (s) & 6.5$\pm$1.0 \\
Duration (until 10\% of peak rate) (s) & 82.5$\pm$1.0 \\
Bolometric peak luminosity ($10^{38}$ erg~s$^{-1}$) & 2.0$\pm$0.5  \\
Bolometric radiated energy ($10^{39}$ erg) & 3.9$\pm$0.2 \\
Persistent L$_X$ (0.5--10~keV; $10^{36}$ erg~s$^{-1}$) & 3.3$\pm$0.1  \\
\hline\hline
\end{tabular}
\end{minipage}
\label{table:burst}
\end{table}

\begin{figure*}
  \begin{center}
  \resizebox{1.5\columnwidth}{!}{\rotatebox{-90}{\includegraphics[]{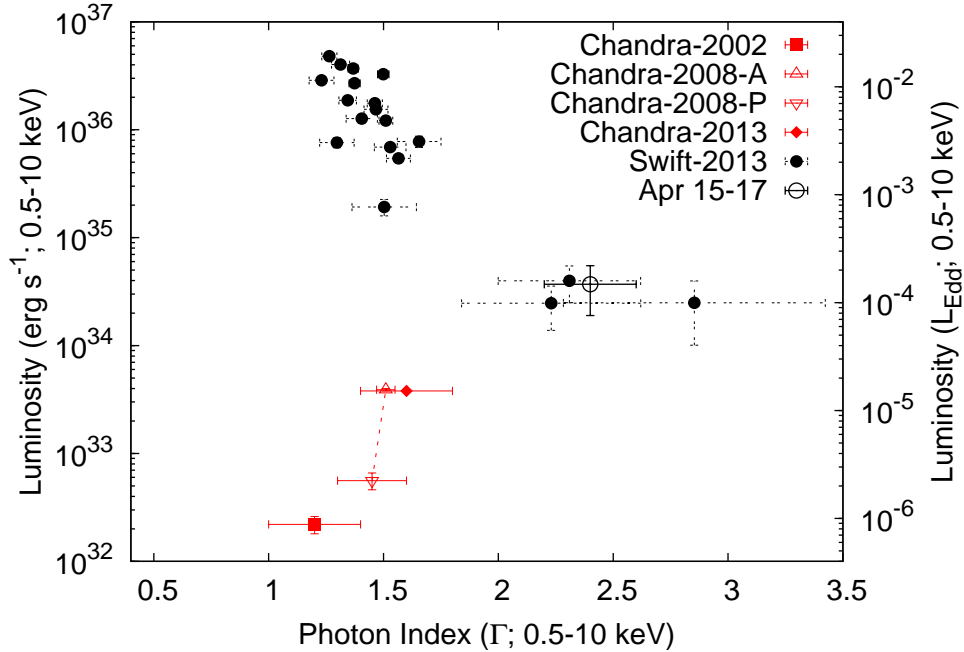}}}
  \caption{Photon index ($\Gamma$) vs. luminosity (L$_X$) for
  IGR~J18245--2452 from outburst to quiescence, spanning five
  orders of magnitude in L$_X$. The $\Gamma - L$ anticorrelation
  (``softening towards quiescence'') at relatively high L$_X$
  ($10^{34} - 10^{37} erg s^{-1}$) turns into a positive correlation
  below L$_X$$\sim 10^{-4} L_\mathrm{Edd}$. The red dotted line
  connects the 2008 active and passive states (Sec.~\ref{sec:qui};
  Figs.~\ref{fig:qspec} \& \ref{fig:qlc}).}
    \label{fig:gamma}
 \end{center}
\end{figure*}

The burst was detected about 10~d after the beginning of the outburst
\citep{Papitto13,Linares13}, when the luminosity and inferred mass
accretion rate ($\dot{M}$) were a few percent of the Eddington limit
($\sim$4\% $L_{Edd}$ from the measured 0.5--10~keV luminosity of
3.3$\times 10^{36}$~erg~s$^{-1}$ and using $f_{bol}$=3). At this
$\dot{M}$, close to the boundary between the pure He and mixed H/He
ignition regimes, H is expected to burn stably and thus He ignition
triggers the bursts \citep{Bildsten98}. The long decay timescale,
together with the lack of radius expansion, suggest that the burst
ignites in a mix of H and He \citep[e.g.,][]{Galloway08}. The total
radiated bolometric burst energy that we measure,
3.9$\times$10$^{39}$~erg, corresponds to an ignition depth of
$\sim$9.6$\times$10$^{7}$~g~cm$^{-2}$ \citep[assuming solar-abundance
homogeneousy-distributed fuel releasing 4.4 MeV~nucleon$^{-1}$;
e.g.,][]{Galloway08}, which can be reached when accreting at 4\% of
the Eddington rate in only $\sim$0.3~d. This crude estimate of the
expected burst recurrence time suggests that more thermonuclear bursts
occurred during the 2013 $\sim$25d-long outburst, but given that {\it
Swift}-XRT observed \IGR{} for a total accumulated exposure of
$\sim$0.4d, it is not surprising that it only detected one
thermonuclear burst.

\subsection{Between outburst and quiescence}
\label{sec:ob2q}

We find an anticorrelation between the persistent luminosity, L$_X$,
and the photon index, $\Gamma$, measured by {\it Swift} during the
outburst peak and decay when L$_X$/L$_{Edd}$ decreased from 10$^{-2}$
to 10$^{-4}$. $\Gamma$ increases from 1.3 above
L$_X$/L$_{Edd}$$=$10$^{-2}$ to $\sim$2.5 at
L$_X$/L$_{Edd}$$=$10$^{-4}$. This softening during the decay to
quiescence (already noticeable in Fig.~\ref{fig:ob}, top two panels)
can be clearly seen in Figure~\ref{fig:gamma}, where we plot L$_X$
vs. $\Gamma$ over five orders of magnitude in L$_X$ (note that
special care was taken to avoid background contamination in the {\it
Swift} spectra; Sec.~\ref{sec:swift}). In order to test the
significance of this anticorrelation, we calculate Spearman's
rank-order coefficient ($r$) using all (19) values of [L$_X$,
$\Gamma$] measured with {\it Swift} and find $r=-0.80$, which deviates
from 0 at the 4.1$\sigma$ confidence level (the same test using only
the brightest {\it Swift}-XRT data at L$_X$$>$10$^{35}$~erg~s$^{-1}$
gives $r=-0.67$ and a 2.8$\sigma$ significance). Such softening of the
X-ray spectrum from (hard state) outburst to quiescence has been
previously seen in both NS and BH transients \citep[][see discussion
in Sec.~\ref{sec:accretion}]{Wu08,Armas-Padilla11,Plotkin13}.

The absorbing column densities during the 2013 outburst
(N$_H$=[3--5]$\times10^{21}$~cm$^{-2}$ at L$_X$$>$10$^{34}$erg~s$^{-1}$)
are higher than those measured in quiescence by {\it Chandra}
(Sec.~\ref{sec:qui}, Table~\ref{table:qspec}).
This can be seen from our fits to the {\it Swift} persistent spectra
(Figure~\ref{fig:ob}), which yield an average
N$_H$=[4.0$\pm$0.1]$\times10^{21}$~cm$^{-2}$, higher than the value of
[2.6$\pm$0.2]$\times10^{21}$~cm$^{-2}$ measured in the 2008
observations when the luminosity was
L$_X$=[0.6--4]$\times$10$^{33}$~erg~s$^{-1}$. Fixing N$_H$ to the
value found with {\it Swift} during outburst does not yield acceptable
spectral fits to the {\it Chandra} 2008 quiescent spectrum (reduced
$\chi^2$ increases from 1.0 to 1.3 for 248 dof). This result suggests
that intrinsic changes in the absorbing material take place between
outburst and quiescence, perhaps linked to the increased mass
accretion rate during outburst, and cautions against the blind use of
N$_H$ values measured in outburst for quiescent studies.

\subsection{Variable quiescence}
\label{sec:qui}

All {\it Chandra} spectra are well fitted with a simple absorbed power
law model. We present in Table~\ref{table:qspec} the results of the
spectral fits to the {\it Chandra}-ACIS quiescent spectra at different
epochs, between 2002 and 2013. We show the resulting L$_X$--$\Gamma$
relation in Figure~\ref{fig:gamma}, which covers the
L$_X$/L$_{Edd}$$=$[10$^{-6}$--10$^{-4}$] range, together with the
{\it Swift} data at higher luminosities,
L$_X$/L$_{Edd}$$=$[10$^{-4}$--10$^{-2}$].

Our {\it Chandra} analysis shows that \IGR{} is a hard, purely
non-thermal X-ray source in quiescence, with a 0.5--10~keV luminosity
that varies by more than one order of magnitude,
2.5$\times$10$^{32}$~erg~s$^{-1}$ in 2002 and up to
3.8$\times$10$^{33}$~erg~s$^{-1}$ in 2008, while the photon index
$\Gamma$ remained between 1 and 1.5. 
Soon after its 2013 outburst, during a {\it Chandra} observation taken
on 2013-04-28 \citep{Homan13}, we detected \IGR{} with the same
luminosity (3.9$\times$10$^{33}$~erg~s$^{-1}$) and spectral index
($\Gamma$$\simeq$1.5) as those measured in 2008 (during its active
state, see below).

We show in Figure~\ref{fig:qspec} a comparison between the unfolded
quiescent spectra from \IGR{} at different epochs, to illustrate the
large luminosity changes that take place with little or no change in
the spectral shape.
Given that all the spectra taken at luminosities below
10$^{34}$~erg~s$^{-1}$ (L$_X$/L$_{Edd}$$<$10$^{-4}$), are harder than
those measured by {\it Swift} in the latest stages of the outburst
decay, we define quiescence as L$_X$/L$_{Edd}$$<$10$^{-4}$
(Figure~\ref{fig:gamma}). With this definition, and extrapolating the
latest flux decay seen by {\it Swift}, we estimate that the outburst
finished around 2013-04-22 (with an uncertainty of about 4 d) and
lasted for 25~d, which implies that both 2013 {\it Chandra}
observations \citep{Homan13,Papitto13b} were taken when \IGR{} was in
quiescence.

The most stringent limits on the thermal NS component come from the
observations where \IGR{} is the faintest, taken in 2002.  We extracted
a spectrum from \IGR{} from the three 2002 ACIS-S observations combined
(Sec.~\ref{sec:chandra}). The spectrum is well fitted with a simple
absorbed power-law model, which yields a 0.5--10~keV luminosity of
[2.5$\pm$0.8]$\times 10^{32}$~erg~s$^{-1}$
(Table~\ref{table:qspec}). In order to get upper limits on the NS
temperature and luminosity in quiescence, we added a NS atmosphere
component to the spectral model (\textsc{NSATMOS}, \citealt{Heinke06},
assuming a 1.4 \Msun, 10 km NS with normalization fixed at 1 and a
distance of 5.5~kpc). We then fitted the spectrum using chi-squared
statistics (binning to 15 counts/bin, which left us with 6 bins) and
using unbinned data and the C-statistic. We obtain consistent
constraints from the two methods. We find upper limits (at 90\%
confidence) on the 0.1--10~keV (gravitationally redshifted) thermal
quiescent luminosity $L_{NS}^{\infty} < 7\times10^{31}$~erg~s$^{-1}$
and on the corresponding (intrinsic, redshift-corrected) NS surface
effective temperature $T_{eff} < 6.6\times10^5$~K.

The two public {\it Chandra} HRC-S observations
(Sec.~\ref{sec:chandra}) show \IGR{} at slightly
fainter levels than the 2002 ACIS observations.
We find net source count rates of
(7.5$\pm$0.9)$\times$10$^{-4}$~c~s$^{-1}$ and
(7.1$\pm$1.0)$\times$10$^{-4}$~c~s$^{-1}$ in the November 2002 and May
2006 observations, respectively.
Although these data provide no spectral information, we can estimate
the quiescent luminosity assuming that the spectral shape is similar
to that measured in the {\it Chandra}-ACIS observations. For
N$_H$=2.6$\times$10$^{21}$ cm$^{-2}$ and $\Gamma$ in the range
1.0--1.5 we obtain L$_X$=[1.1--2.1]$\times10^{32}$~erg~s$^{-1}$ during
the 2002 and 2006 HRC observations.

Interestingly, comparing our {\it Chandra} and {\it Swift} results
shows that the spectrum does not continue to soften at luminosities
L$_X$/L$_{Edd}$$<$10$^{-4}$, but it hardens instead
(Fig.~\ref{fig:gamma}). All four {\it Chandra} spectra (taken in 2002,
2008 and 2013) are consistent with $\Gamma$$\simeq$1.5 (a fit to a
constant $\Gamma$ gives $\chi^2$=2.7 for 3 d.o.f. and a Spearman rank
test yields no significant correlation, although the sample is
obviously small and fractional errors on $\Gamma$ are large). However,
including the three lowest-L$_X$ {\it Swift} measurements (at
L$_X$/L$_{Edd}$$\sim$10$^{-4}$) yields $r=0.93$ and a positive correlation
with a 3$\sigma$ significance. This strongly suggests that the X-ray
spectrum hardens while reaching deep quiescence
(L$_X$/L$_{Edd}$=10$^{-4}$$\rightarrow$10$^{-6}$), after softening in the
hard state during the outburst decay
(L$_X$/L$_{Edd}$=10$^{-2}$$\rightarrow$10$^{-4}$, see
Sec.~\ref{sec:ob2q}). More frequent and sensitive observations of the
decay to quiescence may allow a better measurement of this spectral
transition between outburst and quiescence. We tentatively place the
luminosity at which this transition occurs at
L$_X$/L$_{Edd}$$\sim$10$^{-4}$ (see Figure~\ref{fig:gamma}).

\begin{figure}
  \begin{center}
  \resizebox{1.0\columnwidth}{!}{\rotatebox{-90}{\includegraphics[]{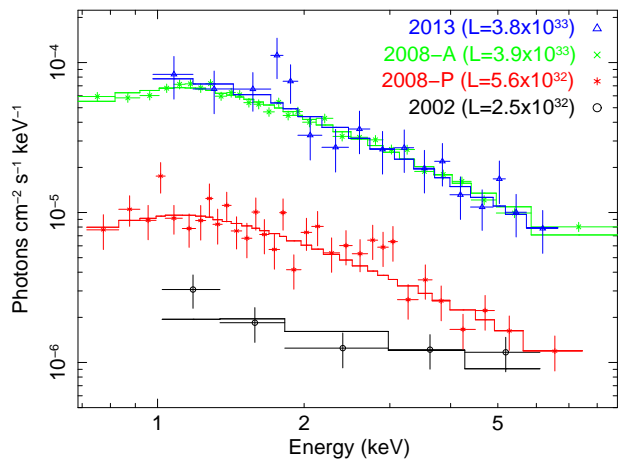}}}
  \caption{
Quiescent {\it Chandra}-ACIS-S unfolded spectra from IGR~J18245--2452
(S23) at different epochs, as noted (A and P stand for active and
passive states within quiescence in 2008, see text). The 0.5--10~keV
luminosity in erg~s$^{-1}$, which varies by more than one order of
magnitude, is indicated between parenthesis.
}
    \label{fig:qspec}
 \end{center}
\end{figure}

\begin{figure*}
  \begin{center}
  \resizebox{2.0\columnwidth}{!}{\rotatebox{-90}{\includegraphics[]{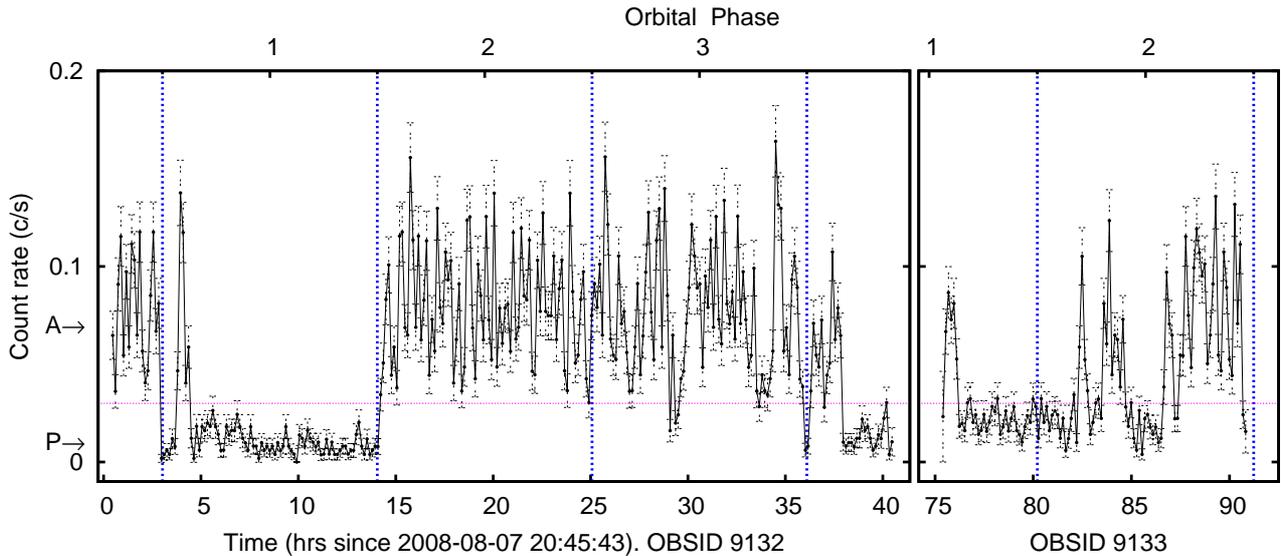}}}
  \caption{Quiescent light curve (500~s bins, background-subtracted)
    from IGR~J18245--2452 (S23), measured with {\it Chandra}-ACIS in
    2008 August 7 (left) and 10 (right). The source switches
    between active (A) and passive (P) states (defined by the
    horizontal purple line), with a factor $\sim$7 change in luminosity
    and no detectable spectral change (Sec.~\ref{sec:qui}). Vertical
    blue lines mark orbital phase 0.5 \citep[from the ephemerids given
    by][]{Papitto13b}.}
    \label{fig:qlc}
 \end{center}
\end{figure*}

The 2008 quiescent {\it Chandra}-ACIS light curves show striking
variability on several time scales (see Figure~\ref{fig:qlc}). We
find two clearly distinguishable states or ``modes'' in the
light curve, at two distinct average count rates, and rapid changes
between them (within less than 500~s).
We identify a low luminosity {\it passive state} and a high luminosity
{\it active state} during quiescence, below and above a {\it
Chandra}-ACIS 0.5--8~keV count rate of 0.03 c~s${-1}$, respectively
(dashed purple line in Figure~\ref{fig:qlc}). These are most clearly
visible in the 144~ksec-long observation taken on 2008-08-07
(observation 9132; Fig.~\ref{fig:qlc}, left), but also three days
later in the 55~ksec-long observation taken on 2008-08-10 (observation
9133; Fig.~\ref{fig:qlc}, right).
Even though it was taken when L$_X$ was similar to the 2008
active state (Table~\ref{table:qspec}), the 2013 {\it Chandra} light
curve does not show rapid intensity changes. However, we cannot rule
out the presence of mode switching during that observation, as the
exposure time was shorter (15~ksec) and the count rate was lower
(1.2$\pm$0.4$\times$10$^{-2}$~c~s$^{-1}$) than those in the 2008
observation.

After $\sim$9~ksec in the active state since the beginning of
observation 9132, the source switched to the passive state for about
3~ksec and after a short ($\sim$2.5~ksec) re-flare to the active
state, it stayed in the passive state for another $\sim$35~ksec
(Fig.~\ref{fig:qlc}, top). Then \IGR{} switched back to the active state
for $\sim$85~ksec, with a few short ($\sim$1~ksec) drops in count rate
suggestive of short excursions to the passive state. For the last
$\sim$9.5~ksec the source stayed in the passive state. Three days
later, observation 9133 starts with a $\sim$2.5~ksec flare to the
active state, then after 38~ksec of passive state with two more short
(about 2 and 4 ksec-long) flares the source switches back to the
active state for about 27 skec, and switches back to the passive state
at the end of the observation.
We also find rapid ($\sim$500--1000~s) and less pronounced (factor
$\sim$2--3 changes in count rate) variability in the light curves of
both the active and passive states.

The hardness ratio (defined as the ratio of counts in the 2--8~keV and
0.5--2~keV bands, and calculated in 2~ksec bins) is consistent with
being constant during both 2008 observations. We extracted spectra of
the active and passive states, using the above definition and the same
procedures explained in Sec.~\ref{sec:chandra}. Both are well-fit with
an absorbed power-law model, and the best-fit $\Gamma$ and $N_H$ in
the active and passive states are fully consistent (see
Table~\ref{table:qspec}).
In order to search for small spectral variations, we repeated the
fits to the active and passive state spectra keeping $N_H$ linked or
fixed to its average value, and again found $\Gamma$ values consistent
within the reduced (1--5\%) fractional errors.
We measure a 0.5--10~keV luminosity of 3.9$\times
10^{33}$~erg~s$^{-1}$ and 5.6$\times 10^{32}$~erg~s$^{-1}$ for the
active and passive state, respectively. Thus when the source switches
between active and passive quiescent states the luminosity changes by
a factor of $\sim$7 with no detectable change in the
spectrum. Figure~\ref{fig:qspec} shows the active and passive state
spectra, as well as the other quiescent spectra of \IGR{} measured with
{\it Chandra} in 2002 and 2013.

It is perhaps worth noting that the first passive state begins and
ends at orbital phase 0.5 \citep[first two vertical lines in
Fig.~\ref{fig:qlc}, from orbital ephemerids reported
by][]{Papitto13b}, i.e., when the NS is at the descending node
(approaching most rapidly). Even though one of the short dips later in
that observation also occurs at orbital phase 0.5, the variability
pattern does not recurr and in general the source switches between
active and passive states at different orbital phases.

\section{Discussion}
\label{sec:discussion}

\subsection{Underluminous accretion flows and softening to quiescence}
\label{sec:accretion}

We found that when \IGR{}'s luminosity decayed from
L$_X$/L$_{Edd}$$\sim$10$^{-2}$ to L$_X$/L$_{Edd}$$\sim$10$^{-4}$ the
{\it Swift}-XRT (0.5--10~keV) spectrum could be described by a simple
power-law, and the photon index $\Gamma$ increased from 1.3 at the
highest luminosities to $\sim$2.5 at the lowest luminosities
(Section~\ref{sec:ob2q}). Similar softening has been seen before in
other systems, both NS \citep{Armas-Padilla11,Degenaar13} and BH
\citep{Wu08,Sobolewska11,Plotkin13} transients, and it appears to be a
general observable of LMXBs in the hard state when they transit to
quiescence. In the context of advection dominated (radiatively
inefficient) accretion flows \citep[ADAFs;][]{Narayan95}, this
behavior can be explained by a change in the optical depth for Compton
up-scattering: as $\dot{M}$ decreases, the optical depth of the ADAF
and the fraction of hard/up-scattered photons decrease, causing a
softening of the X-ray spectrum \citep[][predict a change in $\Gamma$
of 1.5 $\rightarrow$ 2.2 due to this effect]{Esin97}.
Other physical jet-based interpretations for this softening towards
quiescence exist, which assume that X-ray emission from the jet
dominates in quiescence \citep[see, e.g., discussion in Sec.~4.2
of][and references therein]{Plotkin13}.

In the case of accreting NSs the interpretation of the observed
X-ray spectra at low luminosities should also consider potential
thermal surface emission, coming from either low-level accretion onto
the NS or from the crust heated up during outburst.
Indeed, a thermal component has been found in X-ray spectra of
NS-LMXBs at a luminosity of a few times 10$^{34}$ erg~s$^{-1}$, in
both transient \citep{Degenaar13} and persistent accretors
\citep{Armas-Padilla13}. It is unclear at which luminosity and in
which systems this component becomes important, and therefore we
cannot assess whether or not it contributes to the softening of
NS-LMXBs when they decay from L$_X$/L$_{Edd}$$\sim$10$^{-2}$ to
L$_X$/L$_{Edd}$$\sim$10$^{-4}$. The lowest luminosity {\it
Swift}-XRT spectra of \IGR{}, at L$_X$$\simeq$4$\times$10$^{34}$
erg~s$^{-1}$, do not show evidence of a thermal component and allow
only weak constraints on the presence of such component. We derive a
90\% upper limit on the thermal fraction of $<$24\% (defined as the
fractional contribution of a blackbody component to the 0.5--10 keV
luminosity).
This upper limit is derived assuming a temperature of 0.3~keV, like
that measured by \citet{Armas-Padilla13} at similar L$_X$, and is
compatible with the range of values found by the same authors in other
NS-LMXBs (8--53\%).

The deep {\it Chandra} observations allow us to place stringent
constraints on the presence of a thermal component at lower
luminosities, which correspond to a thermal fraction lower than 3\% at
L$_X$$\simeq$2.5$\times$10$^{32}$ erg~s$^{-1}$ (Secs.~\ref{sec:qui}
\&~\ref{sec:nonthermal}). We thus conclude that if an undetected
thermal component is responsible for the spectral softening in the
decay to quiescence observed with {\it Swift} (Sec.~\ref{sec:ob2q}),
then this component should virtually disappear when the luminosity
drops below $\sim$10$^{34}$~erg~s$^{-1}$. This in turn suggests that
such thermal component would be accretion-powered rather than
crustal-heat-powered, and that the accretion flow is able to reach the
NS surface and produce thermal emission down to
L$_X$/L$_{Edd}$$\sim$10$^{-4}$ \citep[see also][]{Armas-Padilla13}. We
note, however, that for dipole magnetic fields of 10$^8$~G accretion
onto the NS surface at such low luminosities may be halted by the
propeller mechanism \citep{Illarionov75,Campana98}.

\subsection{Active and non-thermal quiescence}
\label{sec:nonthermal}

We have shown that in quiescence, which we define as
L$_X$/L$_{Edd}$$<$10$^{-4}$ (Sec.~\ref{sec:qui}), the new NS X-ray
transient in M28 \IGR{} is fully dominated by a hard power-law spectral
component, with a photon index of 1--1.5. By comparing {\it
Chandra}-ACIS observations of M28 taken years apart (between 2002 and
2013), we have also shown that the quiescent 0.5--10~keV luminosity
varies by more than an order of magnitude (between
2.5$\times$10$^{32}$~erg~s$^{-1}$ in 2002 and 3.8$\times$10$^{33}$
erg~s$^{-1}$ in 2008 and 2013).
All these properties resemble the quiescent behavior of the NS X-ray
transient EXO~1745-248 in the globular cluster Terzan 5 in great
detail \citep[][]{Wijnands05c,Degenaar12c}. The reason for such large
fluctuations in quiescence is not clear \citep[see, e.g.,][for an
in-depth discussion]{Degenaar12c}. A possible explanation is that
continued low-level accretion is occurring in the system, although if
that were to reach the NS surface one would expect detectable thermal
emission \citep[e.g.,][]{Zampieri95,Soria11} instead of a purely
non-thermal quiescent spectrum.

On the other hand, the similarities between the X-ray emission of some
quiescent NS-LMXBs and MRPs \citep{Bogdanov05} also apply to
\IGR{}. Namely, non-thermal and variable X-ray emission has also been
seen in the NS-LMXBs Cen~X-4 \citep{Rutledge01}, Aql~X-1
\citep{Campana03b}, SAX~J1808.4--3658 \citep{Campana02} and in the
MRPs 47~Tuc~W \citep{Bogdanov05} and PSR~J1023+0038 \citep{Archibald10}.

We can compare the quiescent X-ray properties of \IGR{} and
PSR~J1023+0038 with some more detail, since these are the only two
MRPs that have shown evidence for accretion. They share low-luminosity
and remarkably hard quiescent spectra, with
L$_X$$\simeq$2.5$\times$10$^{32}$~erg~s$^{-1}$, $\Gamma$$\simeq$1.2
and L$_X$$\simeq$9.4$\times$10$^{31}$~erg~s$^{-1}$,
$\Gamma$$\simeq$1.26 for \IGR{} and PSR~J1023+0038,
respectively. \citet{Archibald10} reported a marginal detection of a
thermal component in PSR~J1023+0038, contributing with 6\% of L$_X$. Our upper
limit on a blackbody component at the lowest L$_X$ ($<$3\%,
Sec.~\ref{sec:qui}) rules out the presence of a similar component in
\IGR{}. The same authors found modulations of the X-ray flux from
PSR~J1023+0038 at the orbital (4.8~hr) and spin (1.7~ms) periods. The data
presented herein do not allow a search for ms X-ray pulsations at low
L$_X$, due to insufficient time resolution. The long-term variability
that we find during the 2008 observations is more rich than that found
in PSR~J1023+0038, and it does not seem to be related with the binary orbit
(Secs.~\ref{sec:qui} \& \ref{sec:pulsar}).

No thermal component is detected at the lowest luminosities
(Sec.~\ref{sec:qui}), corresponding to an upper limit on the
bolometric intrinsic thermal luminosity of 1.3$\times10^{32}$
erg~s$^{-1}$ (for a gravitational redshift of 1.3).
We can compare this to estimates of the incandescent thermal emission
produced by deep crustal heating
\citep{Haensel90,Haensel08,Brown98,Degenaar13}, which should yield a
luminosity $\simeq \langle\dot{M}\rangle Q / m$, where we take $Q \simeq
$2~MeV as the total heat deposited per accreted nucleon,
$\langle\dot{M}\rangle$ is the long-term average mass accretion rate
onto the NS, and $m$ the atomic mass unit. We estimate an average
$\dot{M}\sim 2\times 10^{16}$~g~s$^{-1}$ in the 2013 outburst from our
measured {\it Swift}-XRT luminosities (Sec.~\ref{sec:burst}; using
$f_{bol}$=3 and assuming accretion onto a 1.4\Msun-mass 10~km-radius
NS). For a 25~d outburst duration, this implies an outburst recurrence
time longer than $\sim$20~yr in order to produce a luminosity
lower than our upper limit.
Alternatively, enhanced neutrino cooling may be acting to reduce the
NS core temperature in \IGR{} down to the observed levels ($T_{eff} <
6.6\times10^5$~K, Sec.~\ref{sec:qui}), as reported in other NS
transients \citep[see, e.g.,][]{Heinke07,Wijnands13}.
Given that \IGR{} is active as a rotation-powered pulsar when it is
not accreting \citep[as the MRP PSR~J1824--2452I;][]{Papitto13b}, we
can compare the upper limit on its thermal luminosity with the typical
thermal emission from other MRPs. According to most pulsar models, the
polar caps can be heated up by relativistic particles generated when
the radio pulsar is on, producing thermal pulsed X-ray emission
\citep[e.g.,][]{Zavlin07}. Thermal luminosities measured in MRPs are
typically in the range 10$^{30}$--10$^{31}$~erg~s$^{-1}$, which is
compatible with our upper limit of 7$\times10^{31}$ erg~s$^{-1}$.

When the non-thermal power-law spectral component in quiescent
NS-LMXBs is detected and well constrained, the photon index is
similarly low, between 1 and 1.5 (we note that the error bars are
relatively large). Our results for \IGR{} are consistent with these
typical NS values, which seem to show harder quiescent spectra than
black hole (BH) systems, most of which have a photon index of $\sim$2
in quiescence \citep[][]{Corbel06,Plotkin13}.
Thus, while we find evidence for a spectral hardening below
L$_X$/L$_{Edd}$$\sim$10$^{-4}$ (with the photon index $\Gamma$
reaching 1--1.5; Sec.~\ref{sec:qui}), in BH transients $\Gamma$ seems
to saturate below L$_X$/L$_{Edd}$$\sim$10$^{-5}$ at a value close to 2
\citep{Plotkin13}.
This suggests that different processes cause the quiescent power-law
component in NS and BH X-ray binaries. For BHs this emission is likely
due to some form of radiatively inefficient accretion flow but it is
possible that in the case of NSs another process is at work. Two
emission mechanisms have been proposed for the non-thermal X-rays from
NS-LMXBs in quiescence, in order of increasing luminosity: i) an {\it
intrabinary shock} between the pulsar wind and the flow of mass
transferred/outflowing from the companion star \citep[originally
proposed for MRPs,][]{Arons93,Campana98}; and ii) accretion onto the NS
magnetosphere, or {\it magnetospheric accretion} \citep{Campana98}.

\subsection{At the boundary between accretion- and rotation-powered X-rays?}
\label{sec:pulsar}

We found striking variability in the 2008 {\it Chandra} observations
of \IGR{} (Sec.~\ref{sec:qui}): sharp transitions between two distinct
(active and passive) modes, namely, recurrent, large (factor 7) and
rapid ($\lesssim$500~s) changes in L$_X$, in the [0.6--3.9]$\times
10^{33}$~erg~s$^{-1}$ range. The detected emission was purely
non-thermal and the spectra of the active and passive states were
consistent within the errors (photon index $\simeq$1.5 and no
detectable change in absorption; Table~\ref{table:qspec}).
\citet{Campana08c} found that the AMP SAX~J1808.4--3658 seems to
switch between two states when approaching quiescence. Those states
were apparent only in the long-term (100~d long) {\it Swift} light
curve, they showed clear spectral variability and a smaller change in
luminosity (L$_X$=[1.5--5]$\times 10^{32}$~erg~s$^{-1}$ and
$\Gamma$=[1.7--2.7]). It seems therefore unlikely that they have the
same origin as the active and passive states that we report herein.
The mode switching that we observe in \IGR{}, on the other hand, might
be reminiscent of the correlated X-ray/radio mode switches recently
found by \citet{Hermsen13} in the old isolated pulsar PSR
B0943+10. The latter pulsar is obviously not accreting, and the onset
of accretion in \IGR{}, confirmed at least at high X-ray luminosities,
suggests that the accretion flow plays a role in the state changes
presented in this work, and thus that the physical mechanisms
responsible for mode switches in these two pulsars are different.

Three of the transitions coincide with orbital phase 0.5 (NS at descending
node) but this is not always the case, and in general mode switching
occurs at different orbital phases.
Changes in absorption due to matter in different regions of the binary
system crossing the line of sight would leave an imprint in the
spectrum which we do not observe, and hence cannot explain the mode
changes.
Intrabinary shock emission is expected \citep{Arons93} and possibly
observed in MRPs \citep{Stappers03,Bogdanov05} to be modulated
at the orbital period. Mildly relativistic speeds in the shocked
region can lead to Doppler boosted emission and to orbital modulation
of the measured non-thermal X-rays. In the case of \IGR{}, however, the
large amplitude and aperiodic nature of the observed variability do
not support a scenario where the mode switching is due to Doppler
boosting in the line of sight changing along the orbit.

The available data thus suggest that stochastic changes between two
different non-thermal emission processes (which produce very similar
spectra) are taking place.
One possible explanation is that the observed variability reflects
rapid transitions between magnetospheric accretion at high
luminosities (3.9$\times 10^{33}$~erg~s$^{-1}$) and intrabinary shock
emission at low luminosities (5.6$\times 10^{32}$~erg~s$^{-1}$). In
this scenario, the magnetospheric radius r$_m$ during the 2008 {\it
Chandra} observations would be close to the light cylinder radius
\citep[r$_{lc}$$\simeq$186~km for the 254~Hz spin
frequency;][]{Papitto13b}, and fluctuations in the mass accretion rate
$\dot{M}$ would produce the observed transitions.
A small drop in $\dot{M}$ would cause r$_m$ to expand beyond r$_{lc}$,
which would turn on the radio pulsar and create an intrabinary shock
responsible for the emission in the passive state. An increase in
$\dot{M}$ could bring r$_m$ back inside the light cylinder, turning
off the radio pulsar and allowing for magnetospheric accretion to take
over in the active state.
We stress that this scenario involves a balance between the
pulsar wind and the inner accretion flow, likely in the form of a hot
optically thin flow given the extremely low L$_X$, and that an
optically thick accretion disk may still be present at much larger
radii (r$>>$r$_{lc}$) \citep[see, e.g.,][]{Wang09}.

The active and passive states of \IGR{} correspond to bolometric
luminosities (L$_{bol}$) 1.2$\times$10$^{34}$ erg~s$^{-1}$ and
1.7$\times$10$^{33}$ erg~s$^{-1}$, respectively (using again
$f_{bol}$=3; Sec.~\ref{sec:data}). Assuming a typical neutron star
mass of 1.4 $M_{\odot}$ and radius of 10 km, the magnetospheric radius
$r_{m}$ for spherically symmetric accretion can be expressed as
\citep{Lamb73,Patruno09}:

\begin{equation}
r_{m} =7.8\left[\frac{B}{10^8{\rm\,G}}\right]^{4/7}\left[\frac{L_{bol}}{L_{Edd}}\right]^{-2/7} km
\end{equation} 

Thus for a value of the magnetic field of $B\sim10^{8}$ G
(typical of all AMPs with a measured B field; see, e.g., Table 4 in
\citealt{Patruno12}), the magnetospheric radius $r_{m}$ will move from
about 130~km during the active state up to about 230~km during the
passive state. Given that the light cylinder radius $r_{lc}\simeq186$
km, the magnetosphere will become devoid of material up to the light
cylinder as the accretion rate drops from the active to the passive
state. This in turn might activate the radio pulsar mechanism and turn
on the radio pulsar during the passive states.  A higher value of the
B field, above $10^9$ G, would instead give $r_{m} > r_{lc}$ for the
observed luminosities during the active state. This therefore suggests
that, if this interpretation of the active/passive states is correct,
the B field of \IGR{} is similar to that of other AMPs.

A clear prediction of this simple picture is that the radio pulsar
should be on during the passive state and off during the active
state.
Simultaneous high-time-resolution X-ray and radio observations in
quiescence are thus highly desirable, in order to test this prediction
and shed more light onto \IGR{}'s unique properties.
Measurements of the magnetic field and spin-down rate of
PSR~J1824--2452I will constrain the pulsar wind energy and may allow a
quantitave estimate of the intrabinary shock properties
\citep[e.g.,][]{Bogdanov05}.

Our proposed scenario implies that non-thermal magnetospheric emission
can be quickly (within less than eight minutes) and repeatedly (more
than ten times in three days) quenched and reactivated, and that the
intrabinary shock may form on similar timescales.
Based on the morphology of the X-ray light curves from the MRP PSR~J1023+0038,
\citet{Bogdanov11b} concluded that the intrabinary shock is localized
near or at the surface of the companion star, close to the inner
Lagrangian point.
For IGR J18245-2452, the intra-binary shock during the passive state
is probably much closer to the pulsar, where the pulsar wind runs into
the inner boundary of the accretion disk (given the remarkably short
time scales of switching between the passive and active states, it is
unlikely that the wind blows most of the disk away and then irradiates
the companion). Eventually, the pulsar wind might sweep the disk away
leaving the system in a long-term ``deep quiescence'' state that
resembles the other so-called ``redback'' radio MSP binaries. This is
probably the state IGR J18245-2452 was in during the 2002 and 2006
{\it Chandra} observations.

\footnotetext{After this work was submitted for publication,
supplementary analysis of \IGR{} by \citet{Papitto13b} became
public. Our {\it Chandra} and {\it Swift} results mostly agree, with
two exceptions. Their treatment of additional {\it Swift}-XRT
background (UFB; Sec.~\ref{sec:swift}) seems less conservative (they
claim a last XRT detection on 2013-05-01, while we exclude data taken
after 2013-04-19). They briefly describe part of the {\it Chandra}
2008 light curves as a total quenching of the X-ray emission, while we
have shown that these actually show fast mode switches between active
and passive states (Secs.\ref{sec:qui} \& \ref{sec:pulsar}).}

\textbf{Acknowledgments:}

ML thanks M. Kachelrie{\ss} and the Norwegian U. of Science and
Technology for their hospitality while part of this work was
completed, and J. Casares and M. Roberts for stimulating discussions.
RW is supported by an European Research Council starting grant.
AP aknowledges support from the Netherlands Organization for
Scientific Research(NWO) Vidi fellowship
DA acknowledges support from the Royal Society.
JH and DP acknowledge support by the National Aeronautics and Space
Administration through Chandra Award Number GO3-14032B issued by the
Chandra X-ray Observatory Center, which is operated by the Smithsonian
Astrophysical Observatory for and on behalf of the National
Aeronautics Space Administration under contract NAS8-03060.
This research has made use of data and software provided by the High
Energy Astrophysics Science Archive Research Center (HEASARC). The
scientific results reported in this article are based in part on
observations made by the Chandra X-ray Observatory and data obtained
from the Chandra Data Archive. This work used Swift Gamma-ray Burst
Explorer target-of-opportunity observations (PIs Ferrigno, Romano).


\end{document}